\documentstyle[prl,aps,graphicx]{revtex}

\newcommand{\be}{\begin{equation}}
\newcommand{\ee}{\end{equation}}
\newcommand{\ba}{\begin{array}{l}}
\newcommand{\ea}{\end{array}}
\begin{document} \draft 
\title{Gravitational potential energy of simple bodies:\\
the homogeneous bispherical concavo-convex lens}
\author{Zakir F. Seidov\\
{\it Department of Physics, Ben-Gurion University,
Beer-Sheva 84105, Israel}\\email: seidov@bgumail.bgu.ac.il} \maketitle
\begin{abstract} Gravitational potential energy of a homogeneous
bispherical concavo-convex lens, with the equal absolute values of 
curvature  radii of both surfaces, is studied. Compactness factor 
 notion is introduced and calculated as function of central thickness of
lens. \end{abstract}
\section*{} \subsection*{Introduction}
At present, the potential energy is known analytically for three types of 
{\it homogeneous} self-gravitating bodies: a) ellipsoids [1],
b) concave bispherical lenses [2], and 3) rectangular parallelepipeds [3].
In accompaning paper [4], we present a method of negative density which
allows to obtain the analytical solutions for potential energy of new
kinds of "simple" self-gravitating bodies. Here we present the potential
energy for the homogeneous bispherical concavo-convex lens with radii of
curvature $R_2=-R_1$. Such figure is obtained by cutting out a symmetric
bispheric lens from the sphere (Fig. 1). If we introduce $H$, a central
thickness of lens, then potential energy of the "quasi-symmetric"
bispheric lens is:
\be \ba W={\pi\over 540\,H}\,G\,\rho^2\,R^5\,
 \,\biggl[ 40\,a\,H\,\left( -12 - 16\,{H^2} + {H^4} \right)  + \\
       3\,\pi\,\left( 160 - 240\,{H^2} + 30\,{H^4} - {H^6} \right) +
 960\,\left( -2 + 3\,{H^2} \right) \,
        \arctan {2 - H\over 2\,a} \biggr] ;\ea\ee
here $G$ is constant of Newtonian gravitation, $\rho=constant$ is a
matter density, $R$ is radius of spherical surfaces, $H$ is central
thickness (in units of $R$), and $a=\sqrt{1-H^2/4}$. (Note that in the
caption to Fig. 1 all values are dimensional.) We call
the figure in question "quasi-symmetrical", as in general case, two
spherical surfaces may have  the different absolute values of 
curvature radii. This more general case will be considered elsewhere.
\subsection*{Dependence of potential energy of the lens on central
thickness} The series expansions of $W$ at points $H=0$ and $H=2$ are:
\be \ba W_s(0)/(G\,\rho^2\,R^5)\,=\frac{-64\,H^2\,\pi }{27} +
 \frac{8\,H^4\,\pi }{45} + \frac{H^3\,{\pi}^2}{6};\\
 W_s(2)/(G\,\rho^2\,R^5)\,= \frac{-16\,{\pi }^2}{15} + 
\frac{2\,{\left( -2 + H \right)}^2\,{\pi}^2}{3}.\ea \ee
In general, the function $W(H)$ is monotonic (Fig. 2), simply because the
volume and mass (and so potential energy) of the lens are all increasing
with increasing $H$. \subsection*{Compactness factor} It is of some
interest to consider "compactness" of the lens as function of $H$. We
introduce compactness factor, as characteristic of the potential energy
of the honogeneous bodies, as the dimensionless coefficient :
\be w=-{W\over G\,\rho^2 V^{5/3}};\ee
in general case of inhomogeneous bodies, it is better to use mean
density in the definition of $w$.
As an example, for the honogeneous sphere we have:
\be w_{sphere}={3\over 5}({4\,\pi\over 3})^{1/3}\,=.967195. \ee
This is the maximal possible value of compactness factor for homogeneous
bodies.\\ The volume of the lens is:
\be V=\left( H - \frac{H^3}{12} \right) \,\pi\,R^3.\ee
The dependence of compactness factor $w$ for quasi-symmetric 
concavo-convex lens on central thickness $H$ is shown in  Fig. 3.
The serial expansion of $w$ at point $H=0$ is:
\be w_s(0)=\frac{H^{\frac{1}{3}}\,\left( 128 - 
9\,H\,\pi  \right) }{54\,{\pi}^{\frac{2}{3}}},\ee
which is also shown in Fig. 3. \subsection*{Conclusion}
In conclusion, we present here the analytic formula for gravitational energy of
the homogeneous bispheric concavo-convex lens, when the radii of curvature
of both surfaces of the lens  have the same absolute value. Both potential
energy and compactness factor are shown to be 
monotonic increasing functions of relative central thickness of the lens.
As to applicational aspect of the concavo-convex lens problem, we may
mention the modelling of the solar  system's small bodies  with large
craters, such as Phobos with his large Stickney crater. 

\begin{figure} \includegraphics[scale=.4]{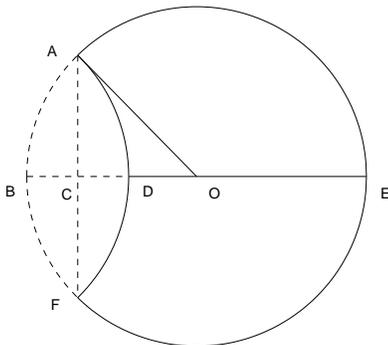}
\caption{"Quasi-symmetric" bispherical concavo-convex lens. Radius of
sphere is $R=OA=OB=OE$; central thickness is $H=DE$; radius, $a$, of base
of additional symmetric bispheric lens ABFD is $a=AC=CF$; half-height,
$h$, or central half-thickness of lens ABFD is  $h=(2\,R-H)/2=R-H/2$;
and $a=\sqrt{R^2-(H+h-R)^2}=\sqrt{R^2-H^2/4}$.} \end{figure}
\begin{figure} \includegraphics[scale=.6]{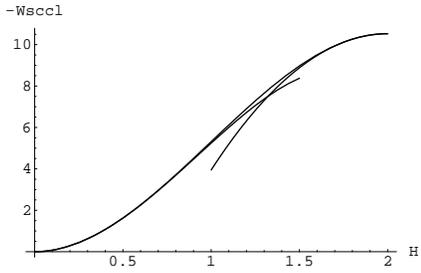}
\caption{Gravitational potential energy, $-W/(G\,\rho_2\,R^5)$, of 
"quasi-symmetric" bispherical concavo-convex lens as function
of dimensionless central thickness $H$. Note that limit $H->0$ corresponds
to "new-born-Moon"-like crescent and limit $H->2$ corresponds to a full
sphere with $-W/(G\,\rho_2\,R^5)=16/15 \,\pi^2$. Also shown are series 
expansions according to Eq. (2).} \end{figure}
\begin{figure} \includegraphics[scale=.6]{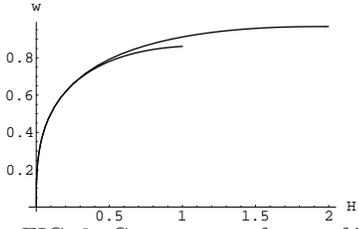}
\caption{Compactness factor of  "quasi-symmetric" bispherical
concavo-convex lens as function of dimensionless central thickness $H$,
according to Eqs. (1), (3) and (5).Also shown is the  series  expansion
according to Eq. (6).} \end{figure} \end{document}